\documentclass[sigconf, nonacm]{acmart}
\AtBeginDocument{%
  \providecommand\BibTeX{{%
    \normalfont B\kern-0.5em{\scshape i\kern-0.25em b}\kern-0.8em\TeX}}}

\setcopyright{acmcopyright}
\copyrightyear{2022}
\acmYear{2022}
\acmDOI{XXXXXXX.XXXXXXX}

\acmConference[Chinese CHI '22]{Chinese CHI '22: The Tenth International Symposium of Chinese CHI (Chinese CHI 2022)}{October 23--October 26, 2022}{Guangzhou, China}
\acmBooktitle{Chinese CHI '22: The Tenth International Symposium of Chinese CHI (Chinese CHI 2022),
  October 22--October 23, 2022, Guangzhou, China}
\acmPrice{15.00}
\acmISBN{978-1-4503-XXXX-X/18/06}

\usepackage{enumerate}
\usepackage{float}
\usepackage{longtable}
\usepackage{booktabs}
\usepackage{supertabular}
\usepackage{array}
\usepackage{multirow}
\usepackage{graphicx}
\usepackage{verbatim}
\usepackage{caption}
\usepackage{colortbl}
\usepackage{url}

\begin{document}
\begin{sloppypar}
\title[A Systematic Review of Meditation, Mindfulness and Virtual Reality]{Reducing Stress and Anxiety in the Metaverse: A Systematic Review of Meditation, Mindfulness and Virtual Reality}
\author{Xian Wang}
\authornote{Both authors contributed equally to this research.}
\email{xian.wang@connect.ust.hk}
\orcid{1234-5678-9012}
\author{Xiaoyu Mo}
\authornotemark[1]
\email{xmoac@connect.ust.hk}
\affiliation{%
  \institution{Hong Kong University of Science and Technology}
  \city{Hong Kong}
  \country{China}
  \postcode{43017-6221}
}

\author{Mingming Fan}
\email{mingmingfan@ust.hk}
\affiliation{
  \institution{Hong Kong University of Science and Technology (Guangzhou)}
  \city{Guangzhou}
  \country{China}
}
\affiliation{
  \institution{Hong Kong University of Science and Technology}
  \city{Hong Kong SAR}
  \country{China}
}

\author{Lik-Hang Lee}
\email{likhang.lee@kaist.ac.kr}
\affiliation{%
  \institution{KAIST}
  \city{Daejeon}
  \country{South Korea}
}

\author{Bertram E. Shi}
\email{eebert@ust.hk}
\affiliation{%
 \institution{Hong Kong University of Science and Technology}
 \city{Hong Kong}
 \country{China}}

\author{Pan Hui}
\email{panhui@ust.hk}
\affiliation{
  \institution{Hong Kong University of Science and Technology (Guangzhou)}
  \city{Guangzhou}
  \country{China}
}
\affiliation{
  \institution{Hong Kong University of Science and Technology}
  \city{Hong Kong SAR}
  \country{China}
}

\renewcommand{\shortauthors}{Wang and Mo, et al.}

\begin{abstract}
 Meditation, or mindfulness, is widely used to improve mental health. With the emergence of Virtual Reality technology, many studies have provided evidence that meditation with VR can bring health benefits. However, to our knowledge, there are no guidelines and comprehensive reviews in the literature on how to conduct such research in virtual reality. In order to understand the role of VR technology in meditation and future research opportunities, we conducted a systematic literature review in the IEEE and ACM databases. Our process yielded 19 eligible papers and we conducted a structured analysis. We understand the state-of-art of meditation type, design consideration and VR and technology through these papers and conclude research opportunities and challenges for the future. 
\end{abstract}

\ccsdesc[500]{Human-centered computing~Virtual reality}

\keywords{Meditation, Mindfulness, Virtual Reality, Literature Review, Interaction, Metaverse}



\maketitle

\section{Introduction}
Meditation, or mindfulness, is grounded in Buddhism, which means meditation in a seated position, and Chinese ``Chan Xiu''. After it was introduced into China, Taoism and Confucianism also adopted it as a cultivation method~\cite{cameron1997openings}. In the 1970s, physiologist Benson found that two transcendental meditation practitioners could control their physiological performance through meditation. Based on this finding, he proposed the relaxation response, which means when a person starts up their ``mental device'' (focusing on a sound or a word or a phrase with a passive attitude), the person will feel relaxed, manifesting in better physiological indicators~\cite{dienstfrey1991mind}. Then Ellen Langer introduced mindfulness into the psychology area. With Dr. Jon Kabat-Zinn coining the term mindfulness-based stress reduction (MBSR), meditation, or mindfulness, is widely used as a cognitive therapy approach in practice~\cite{kabat1995mindfulness}. Many studies have provided evidence that MBSR has benefits for people's stable mental status and overall human well-being~\cite{hempel2015evidence}. 

Virtual Reality (VR) technology can create a virtual world for people to explore without latency or viewing perspective limitations. With the VR head-mounted display, users can enjoy an immersive experience ~\cite{boyd2019introduction}. In addition, 3D modeling software allows artists to create and manipulate the environment in many ways. Users can explore different environments that they cannot visit in real life due to time or space limitations. 
According to Google Trends search, the Metaverse is one of the hottest technology terms in 2021~\cite{usmani2022future}. The Metaverse is defined as a virtual environment that blends physical and digital universes. In particular, VR is an essential technology for the Metaverse~\cite{lee2021all}. 
This realistic 3D virtual environment resembles the actual world, with each user having their own avatar to live life in the virtual world as if it were the real world~\cite{lee2021all,ning2021survey}. 
Because of this, as in the real world, users' levels of stress and anxiety might possibly be alleviated in many virtual realms. Mindfulness and meditation are popular approaches for lowering people's stress and anxiety~\cite{joyce2010exploring,gal2021efficacy}. VR technology or the Metaverse will be able to deliver psychological counseling and therapy services~\cite{jerdan2018head,ifdil2022virtual}.

Theoretically, the attributes of the VR experience are suitable for users to get away from tired daily routines and relax and then enter into a meditation state. Many studies have also indicated that meditation in VR brings various health benefits ~\cite{brown2020virtual, ahmadpour2019virtual,sekula2022virtual,mistry2020meditating,prince2021yoga}. This review aims to understand the latest state-of-art in this area and how VR technology can assist with specific mindfulness tasks. Specifically, we aim to answer the following research questions (RQs):
\begin{itemize}
    \item \textbf{RQ1:} What kinds of meditation or specific mindfulness tasks are involved in the current VR meditation prototype?
    \item \textbf{RQ2:} What factors or variables, corresponding to the meditation type or specific task, can be improved or assisted by Virtual Reality and relevant technologies?
    \item \textbf{RQ3:} How do VR and relevant technologies assist meditation or mindfulness?
\end{itemize}

To answer the RQs, we conducted a systematic literature review  by following the widely used PRISMA method~\cite{moher2009preferred}. Firstly, we identified related records by searching with keywords in the ACM and IEEE databases. By examining the title and abstract, 1,018 records were screened for irrelevant content, and 37 records remained for the next steps. Next, we examined each article's title, abstract, and full text against inclusion criteria. Finally, 18 articles were excluded for ineligibility, and 19 were included for the final analysis to answer the research questions. Regarding RQ1, the result indicated that slightly more than half of the articles didn't have a detailed description of the specific meditation type or mindfulness tasks. Meditation sessions were referred to as mindfulness practice, general meditation, relaxed breath practice, etc. Other mindfulness tasks are mindful breathing, attention focus, body scan, compassion meditation, etc. Regarding RQ2, all articles discussed strengthening the presence of uses via VR visual immersion, while several studies went above and beyond presence by enhancing specific factors that impact meditation, for example, strengthening the awareness of breathing. Finally, regarding RQ3, VR technology makes the user feel immersive by using a head-mounted display (HMD) to render some natural environments (water, vegetation, wind and fire) and some virtual objects (clouds, jellyfish, etc.) that are associated with the natural environment. Some physiological signal detection devices (e.g., breath detection) are also used to interact with virtual objects or to detect the user's meditation state.


\section{Related Work}

In recent years, there has been an increasing amount of literature on meditation and mindfulness based on VR or immersive technology~\cite{terzimehic2019review, dollinger2021challenges, failla2022mediating, sliwinski2017review, arpaia2021narrative,zhang2021effectiveness,o2022virtual,riches2021virtual,kitson2018immersive}, but the vast majority of surveys have focused on how VR technology can help with meditation or mindfulness, rather than on VR technology itself, such as how to design an effective meditation VR program. This section summarizes these studies.

The reviews by Dollinger et al., Failla et al., and Arpaia et al.~\cite{dollinger2021challenges,failla2022mediating,arpaia2021narrative} collectively provide important insights into how VR technology can be used to improve the practice of mindfulness meditation. Together, these studies suggest that VR-based mindfulness interventions can be effective in improving emotional states including anxiety, anger, depression, and tension, and because VR allows precise control of the auditory, visual, and even haptic and olfactory senses of the environment, multisensory immersive VR is more flexible than traditional meditation approaches. However, mainstream VR meditation applications are still based on sound guidance and have little active user input (user interaction with the virtual environment relies primarily on biosignals, such as breathing and heartbeat), therefore Dollinger et al.~\cite{dollinger2021challenges} argue that the full potential of VR technology has not been fully realized at this stage of research. Although the classification of virtual environments and multiple sensory modalities by Dollinger et al. provides much information for this paper, their article does not focus on VR technology. In contrast, our work is more focused on VR technology, and the field of human-computer interaction (HCI). The other two studies do not summarize the design of VR, but only verify that VR technology has its therapeutic effects.

In addition to the mindfulness intervention itself, mindfulness has proven effective for psychological, physical, chronic pain, relaxation, and positive change. Studies by Zhang et al., O et al., Riches et al., and Kitson et al.~\cite{zhang2021effectiveness,o2022virtual,riches2021virtual,kitson2018immersive} have shown that VR technology can also help healthy or clinical groups by practicing mindfulness and thus improving the above symptoms. Because VR is a new technology with a strong sense of immersion and largely reduced real-world distractions, clinical patients would be more receptive to VR-based treatments, which would reduce their stigma~\cite{zhang2021effectiveness}, and general populations could be relaxed by using VR to enter natural virtual environments in workplace settings~\cite{riches2021virtual}. These early studies add to our understanding of VR-supported meditation training, but their research focuses only on the effectiveness of VR technology in healthcare, demonstrating that VR technology can be used as a potential medical tool, but does not provide systematic guidance to future VR program designers on how to design VR mindfulness practice programs. We, as researchers in the field of HCI, are strongly motivated to propose a systematic design framework for VR-based meditation programs aimed at ameliorating the negative effects of social stress that people may face in the post-pandemic era~\cite{riches2021virtual}.

The reviews by Terzimehic et al. and Sliwinski et al.~\cite{terzimehic2019review,sliwinski2017review} although not primarily focused on VR technology, the HCI techniques and interaction technologies involved in both of their work in mindfulness still provide information for this paper. Many of the earlier studies involved sensor technologies to provide biological and neural feedback to participants, but most sensors were only used to record participants' physical states and not for interaction, whereas these possible sensor-based HCI technologies could provide game-like interactions, which were identified by Sliwinski et al.~\cite{sliwinski2017review} as having great potential to improve the mindfulness training process. This coincides with the view of Dollinger et al.~\cite{dollinger2021challenges} that the current stage of research on VR-based mindfulness does not fully exploit the full possibilities of VR technology, which has motivated our current systematic literature review.

\section{Methodology}

\subsection{Search \& Paper Set Extraction}
\begin{figure}[h]
  \centering
  \includegraphics[width=1\linewidth]{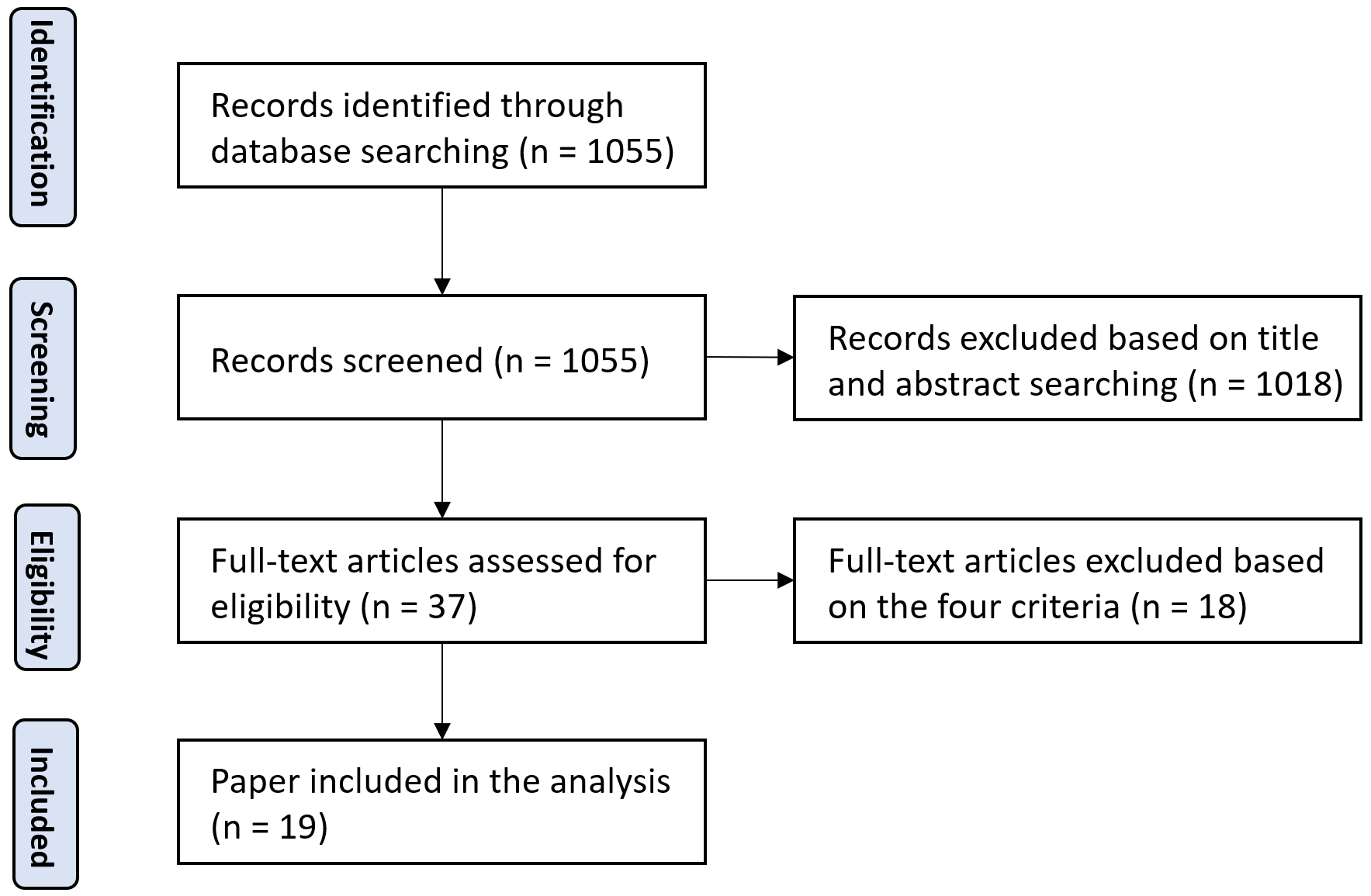}
  \caption{Our literature search process, refer to the PRISMA procedure.}
  \label{fig:Prisma}
\end{figure}

The purpose of this paper is to analyze the current state of research on meditation or mindfulness practice in VR. By analyzing these studies, we aim to describe the assessment of how meditation is currently performed in VR. Based on this description, we will discuss the kinds of meditation that can be integrated with VR technology, meditation design considerations, the VR, or related technologies involved, how they can be implemented in VR and 
the research questions that are under-explored. We followed the four-stage process of the PRISMA method~\cite{moher2009preferred} to conduct a structured review of the existing literature. Figure~\ref{fig:Prisma} shows the four phases of our review process. Our review included papers that met the following four criteria:

  \begin{itemize}
      \item[a)] \textbf{VR technology.} Papers need to address immersive VR technology such as head-mounted displays (HMDs) or CAVEs.
      \item[b)] \textbf{Meditation or mindfulness.} The paper needs to indicate the type of meditation or mindfulness used or the methods commonly used for meditation such as relaxation and concentration.
      \item[c)] \textbf{User experiments.} The paper needs to report a user study and related research data.
      \item[d)] \textbf{Topics.} The paper needs to include meditation or mindfulness in VR as one of the main subjects of study.
  \end{itemize}

We chose IEEE Xplore and ACM Digital Library as our databases for searching and screening, both of which collect a large number of papers published in the fields of VR and HCI. To focus our search on meditation and VR technologies, we selected ``meditation'', ``mindfulness'', ``virtual'', and ``VR'' as keywords and combinations of them. To search more precisely, we applied the advanced search function of the two database websites for Boolean commands (see Tab.~\ref{tab:instructions}). We chose ``virtual'' as a search term to include ``virtual reality'', ``virtual environment'', ``virtual object'' and other related technical terms that would be involved in VR technology. To focus more on the latest technology developments, we only contain papers published within the last 20 years, from 2002 to 2022. To ensure we effectively focus on solid research findings, we only collect peer-reviewed ``research articles'', not including editors, reviews, posters, abstracts, theses, etc. In addition, we only collect papers published in English. A total of 1055 results were obtained, and we collated the titles and abstracts of these publications for the second stage of screening.

\begin{table*}[h]
  \caption{Boolean instructions for IEEE Xplore and ACM Digital Library}
    \renewcommand\arraystretch{1.5}
  \label{tab:instructions}
 \resizebox{\linewidth}{!}{ 
  \begin{tabular}{p{100pt}p{400pt}}
    \toprule
    Database & Boolean instructions \\
    \hline
    IEEE Xplore & ("Document Title": "meditation" OR "Document Title": "mindfulness")  AND ("Document Title": "virtual" OR "Document Title": "VR") OR ("Abstract": "meditation" OR "Abstract": "mindfulness")  AND ("Abstract": "virtual" OR "Abstract": "VR")\\
    ACM Digital Library & Title: ((meditation OR mindfulness) AND (virtual OR VR)) OR Abstract: ((meditation OR mindfulness) AND (virtual OR VR))\\
    \bottomrule
\end{tabular}}
\end{table*}

In the second stage, we used the above inclusion criteria to screen the titles and abstracts of 1,055 papers collected in the first stage.  Two researchers read and annotated the papers separately and then discussed their annotations. If there was a conflict about whether a paper is qualified to be included, they discussed it to gain a consensus. If they could not reach an agreement, a third researcher joined the discussion to jointly resolve the conflicts.

In the third stage, we screened full-text articles for eligibility based on four criteria. The reasons for exclusion at this stage are: (a) although both VR and meditation are involved, meditation is not the main object of the study (Criteria 4); (b) not include user studies' data (Criteria 3); (c) only virtual technology is involved, not adopt the real-time rendering immersive VR (Criteria 1).

\subsection{Analysis \& Coding}
\begin{figure}[h]
  \centering
  \includegraphics[width=1\linewidth]{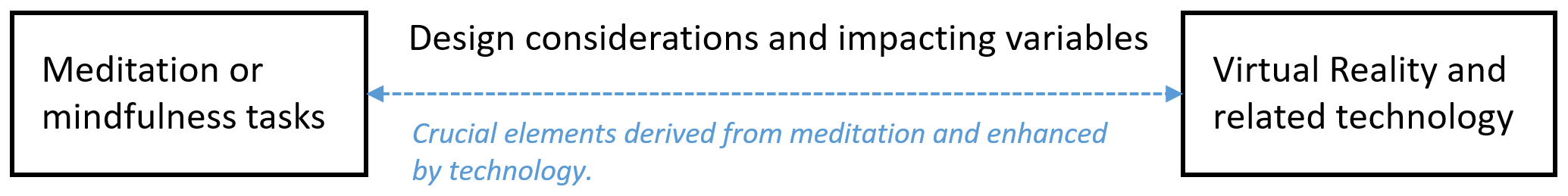}
  \caption{Coding structure, driven by research questions.}
  \label{fig:Coding structure diagram}
\end{figure}

Our coding structure is driven by research questions (see Figure~\ref{fig:Coding structure diagram}). We proposed a coding structure containing three themes. The first theme is the type of meditation and the specific mindfulness tasks. For example, does the study adopt respiration-based meditation or attention-focus-based meditation? 

The second theme is design considerations or impacting variables, which are crucial elements derived from meditation and enhanced by technology. The design consideration means the crucial elements impacting the meditation experience. Prototype design-oriented papers will contain one or more design considerations to promote the user's meditation experience. For example, a prototype will improve the user's awareness of breath through visual cues and biofeedback in the virtual environment. The awareness of breath is the primary design consideration. Impacting variables mean the variables that an empirical study aims to explore their correlation with meditation experience, for example, are nature-based environments positively or negatively correlated with meditation experience. One paper will have both if it proposes a prototype and examines some variables. 

The third theme is VR and related technology. In this theme, the first sub-theme is VR technology. This review refers to the framework proposed in Farhah Amaliya Zaharuddin's team's work as the guide to understanding VR~\cite{zaharuddin2019virtual}. They define VR as \emph{virtual environments, virtual objects, virtual others, and virtual self-representation.} In addition, we also care about the hardware aspect. Therefore, our final structure of the third theme is \emph{VR and Interactive Apparatus, Virtual Environments, Virtual Objects, and Virtual Self-Representation.} Another sub-theme is the detection of physiological signals, which is widely used in many other meditation or mindfulness studies. 
The first theme highlights meditation, and the third theme highlights VR technology, while the second theme is a bridge connecting mindfulness and technology.  

\section{Findings} 

\subsection{Meditation type and mindfulness task (RQ1)}
Slightly more than 
half (10/19) of the articles didn't have a detailed description of the specific meditation type or mindfulness tasks. In these articles, the meditation were referred as mindfulness practice~\cite{roo2017inner}, mindfulness VR experience ~\cite{min2020effects}, general meditation ~\cite{lai2021transcent}, relax breath practice ~\cite{mevleviouglu2021visual}, audio-guided meditation, instructed meditation ~\cite{andersen2017preliminary}~\cite{kazzi2018effects}, calm and mindful experience ~\cite{paredes2018driving}, meditation sitting besides a campfire ~\cite{lee2021study}. 
In Re:Feel~\cite{madzin2021re}, the system is designed to promote Asmaul Husna reciting, while there is a lack of details on how reciting is combined with meditation. In ZenVR~\cite{feinberg2022zenvr}, the curriculum covers a series of meditation topics covering posture, health, and attention as well as introducing different meditation techniques.
There were insufficient illustrations of meditation techniques. The majority of research that has expanded on the particular mindfulness task has focused on mindful breathing~\cite{stepanova2020jel, prpa2018attending,patibanda2017life,yildirim2020efficacy}, i.e., the users feel their breath. Among them, the work of Caglar Yildirim and Tara O’Grady~\cite{yildirim2020efficacy} integrates paying attention to breathing and experiencing the present moment without judgement simultaneously.
Other meditation types or techniques mentioned in the articles are \emph{Kasina}~\cite{moseley2017deep} (a Buddhist technique in Visuddhimagga scriptures, which means one uses some physical objects for concentration before moving to purely mental creations~\cite{dhammaratana2011guide}), \emph{Stability Meditation}~\cite{heeter2020interoceptive} (focus on the feeling of stability by breath, movement, attention, and meditation), \emph{attention focus and body scan}~\cite{kosunen2016relaworld} (focus the attention by gazing at highlighting balls and feel the parts of the body at the present moment), and \emph{compassion meditation} ~\cite{jarvela2021augmented} (focus on the empathetic feelings towards others). In the work of Edirisooriya et al.~\cite{edirisooriya2019smartphone}, deep muscle relaxation and a visual imagery activity are illustrated.

\subsection{Design considerations and impacting variables (RQ2)}

The majority of articles include at least one design consideration, which are crucial components drawn from meditation and augmented by technology.
Although all articles discussed strengthening the presence of users via VR immersion, several studies went above and beyond presence. One-third of the research focused only on presence and immersion.
They upgraded the presence and immersion by creating multi-sensory experiences~\cite{lai2021transcent}, providing multi-modal feedback (haptic and vibration)~\cite{min2020effects}, presenting an uncommon visual experience~\cite{paredes2018driving}, allowing a series of activities~\cite{lee2021study}, providing a natural VR experience with sound~\cite{andersen2017preliminary}, presenting virtual objects for concentration~\cite{moseley2017deep}, and providing synchronized visual and audio guidance~\cite{yildirim2020efficacy}.

In addition to the essential presence and immersion characteristics, several articles strengthen other key aspects of the meditation experience. Several prototypes highlighted the need for breath awareness~\cite{mevleviouglu2021visual,prpa2018attending,patibanda2017life}.
Several articles enhanced more abstract and mind-conscious factors, such as interoception and empathy~\cite{jarvela2021augmented}, a sense of connection in an individual to the world~\cite{stepanova2020jel}, and interoceptive awareness~\cite{heeter2020interoceptive}. 
In RelaWorld, the system presents visual cues to aid attention regulation~\cite{kosunen2016relaworld}. Meanwhile, InnerGarden is the only study that mentioned the design consideration to cultivate no-judgment, acceptance, and autonomy~\cite{roo2017inner}. In ZenVR~\cite{feinberg2022zenvr}, the system contains systematic meditation curriculum, covering a virtual instructor and well-designed knowledge, and compatible environments. The work of Yasitha L. et al.~\cite{edirisooriya2019smartphone} aims to reduce user stress efficiently, based on the user's states by integrating VR technologies, reinforcement learning, and wearable Arduino components.

{\footnotesize
\begin{table*}[h]
\renewcommand\arraystretch{1.5}
  \caption{A summary of the meditation type and design considerations}
  \label{tab:design}
 \resizebox{\linewidth}{!}{ 
  \begin{tabular}{p{90pt}p{120pt}p{270pt}}
    \toprule
    \rowcolor[gray]{0.9} Article & Mindfulness type & Design considerations or impacting variables\\
    \hline
    Min et al.~\cite{min2020effects} & Not specified:\par general mindfulness VR experience session & Design consideration: a sense of presence;\par
Impacting variables: calm content \emph{vs.} disturbing content
\\
\hline
Jarvela et al.~\cite{jarvela2021augmented} & Loving-kindness and compassion meditation & Design consideration: empathy and interception \par
Impacting variables: \par 
dyadic \emph{vs.} solo meditation;\par 
brainwave visualization \emph{vs.} breathing visualization \emph{vs.} both \emph{vs.} no feedback
\\
\hline
Lai et al.~\cite{lai2021transcent} & Not specified: \par general meditation & Design consideration: a hybrid composition of sensory experiences\\
\hline
Stepanova et al.~\cite{stepanova2020jel} & Mindfulness breath & Design consideration: 
a profound sense of connection in an individual to the whole world through a self-transcendent experience; Intimate connection; Capacity to care for the environment
\\
\hline
Kosunen et al.~\cite{kosunen2016relaworld} & Focused attention; body scan & Design consideration: attention regulation; \\
\hline
Mevleviouglu et al.~\cite{mevleviouglu2021visual} & Not specified: \par general breath relaxation practice & Design consideration: Awareness of respiration \\
\hline
Andersen et al.~\cite{andersen2017preliminary} & Not specified: \par An audio guide to meditation & Impacting variables:
An immersive VR experience vs. non-VR experience
\\
\hline
Kazzi et al.~\cite{kazzi2018effects} & Not specified: \par  an instructed meditation & Impacting variables: \par 
HMD 3D visual and audio \emph{vs.} 2D on mobile phone 
Meditation after physical stress \emph{vs.} after mental stress 
\\
\hline
Paredes et al.~\cite{paredes2018driving} & Not specified: \par  a calm and mindful experience & Design consideration: 
Engagement and relaxation \par 
Impacting variables: \par 
A static VR content \emph{vs.} a dynamic VR content \par 
A static car \emph{vs.} a moving car
\\
\hline
Prpa et al.~\cite{prpa2018attending} & Breath  & Design consideration: 
Awareness of breath 
\\
\hline
Madzin et al.~\cite{madzin2021re} & Not specified: \par Asmaul Husna recite & Design consideration: Reproduce the Asmaul Husna recite scenario on the self-help application
\\
\hline
Lee et al.~\cite{lee2021study} & Not specified: \par meditation in a campfire setting with a forest context & Design consideration: 
Active immersion and concentration 
\\
\hline
Ralph Moseley ~\cite{moseley2017deep} & Kasina & Design consideration:
Deep Concentration on the virtual objects (the product of Kasina) in the virtual environment \par 
Impacting variables:
No technical assistant \emph{vs.} associated audio \emph{vs.} entrainment and biofeedback loop \emph{vs.} Virtual Mind Machine
\\
\hline
Roo et al.~\cite{roo2017inner} & Not specified: \par  mindfulness practice & Design consideration:
Suitable guidance, minimalist, non-judgement, promoting acceptance, promoting autonomy, Tangible interaction, choosing the right reality \\
\hline
Patibanda et al.~\cite{patibanda2017life} & Breath  & Design consideration:
Self-awareness of breathing and body
Focused immersion 
\\
\hline
Heeter et al.~\cite{heeter2020interoceptive} & A yoga-based meditation focusing on stability & Design consideration: Interoceptive awareness \par 
Impacting variables: \par 
A yoga-based meditation focusing on stability \emph{vs.} a parallel relaxation mind wandering experience \par 
Eyes closed \emph{vs.} a 96-degree field of view VR headset experience \emph{vs.} a 110-degree field of view VR headset experience 
\\
\hline
Feinberg et al.~\cite{feinberg2022zenvr} & Not specified: \par A series of lesson topics covering posture, breath, and attention as well as different techniques. & Design consideration: learner-oriented experience containing a well-designed curriculum and virtual environments
\\
\hline
Edirisooriya et al.~\cite{edirisooriya2019smartphone} & Deep muscle relaxation for lower stress group and visual imagery activities for moderate or high-stress group & Design consideration: to improve the awareness of presence, also known as mindfulness, by VR environment stimulations, Arduino-based hardware components, and a rewarding mechanism provision.
\\
\hline
Caglar Yildirim and Tara O’Grady ~\cite{yildirim2020efficacy} & Paying attention to breathing and trying to experience the present moment without judgment. & Impacting variables: the immersive visual experience in the virtual environment.
\\

    \bottomrule
\end{tabular}}
\end{table*}
}

\subsection{Virtual Reality and related technology (RQ3)}

\subsubsection{Virtual Reality (VR)}
This section describes all the techniques used in the reviewed articles, divided into VR-related techniques (apparatus, virtual environments, virtual objects, virtual self-representation, etc.) and detection of physiological feedback-related techniques (EEG detection, etc.). We will discuss the technologies involved.

\begin{figure*}[h]
  \centering
  \includegraphics[width=.7\linewidth]{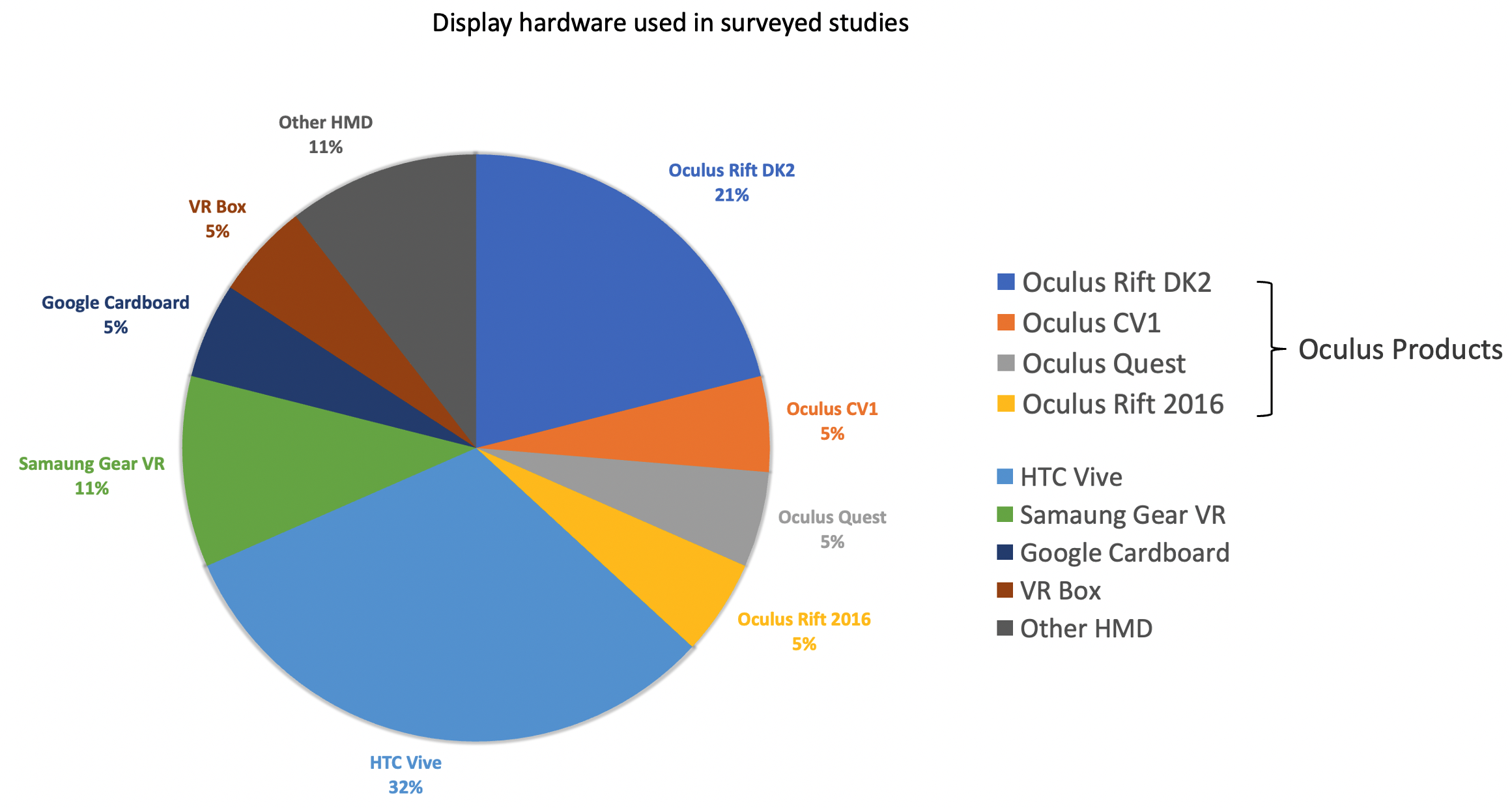}
  \caption{The VR headset models used in the reviewed articles.}
  \label{fig:Devies}
\end{figure*}

~\textbf{VR and Interactive Apparatus.}
Over the past 20 years, all of the work reviewed used head-mounted displays (HMD) as the hardware device for rendering output from virtual environments (see Figure~\ref{fig:Devies}). 
Except for the research conducted by Patibanda et al.~\cite{patibanda2017life} and Yasitha L. et al.~\cite{edirisooriya2019smartphone}, no HMD model was specified. Seven works used Oculus products (four of them used Oculus Rift DK2~\cite{kosunen2016relaworld,roo2017inner,prpa2018attending,paredes2018driving}, one used Oculus CV1~\cite{min2020effects}, one used Oculus Quest~\cite{feinberg2022zenvr} and one used Oculus Rift 2016~\cite{jarvela2021augmented}). Six works used HTC Vive~\cite{stepanova2020jel,mevleviouglu2021visual,lai2021transcent,lee2021study,heeter2020interoceptive,yildirim2020efficacy}, and two used Samsung Gear VR~\cite{andersen2017preliminary,kazzi2018effects}. Two works used inexpensive headsets to cooperate with smartphones (Google Cardboard~\cite{moseley2017deep} and VR Box~\cite{madzin2021re}).

The majority of research used HMD companion controllers as input devices for engaging with the virtual world; however, Roo et al.~\cite{roo2017inner} employed Kinect and Leap motion to detect participant movement. Numerous research (7/19) use physiological signals as part of the interaction with the virtual world; the corresponding devices will be detailed in detail in the next paragraph.

\begin{figure}[h]
  \centering
  \includegraphics[width=1\linewidth]{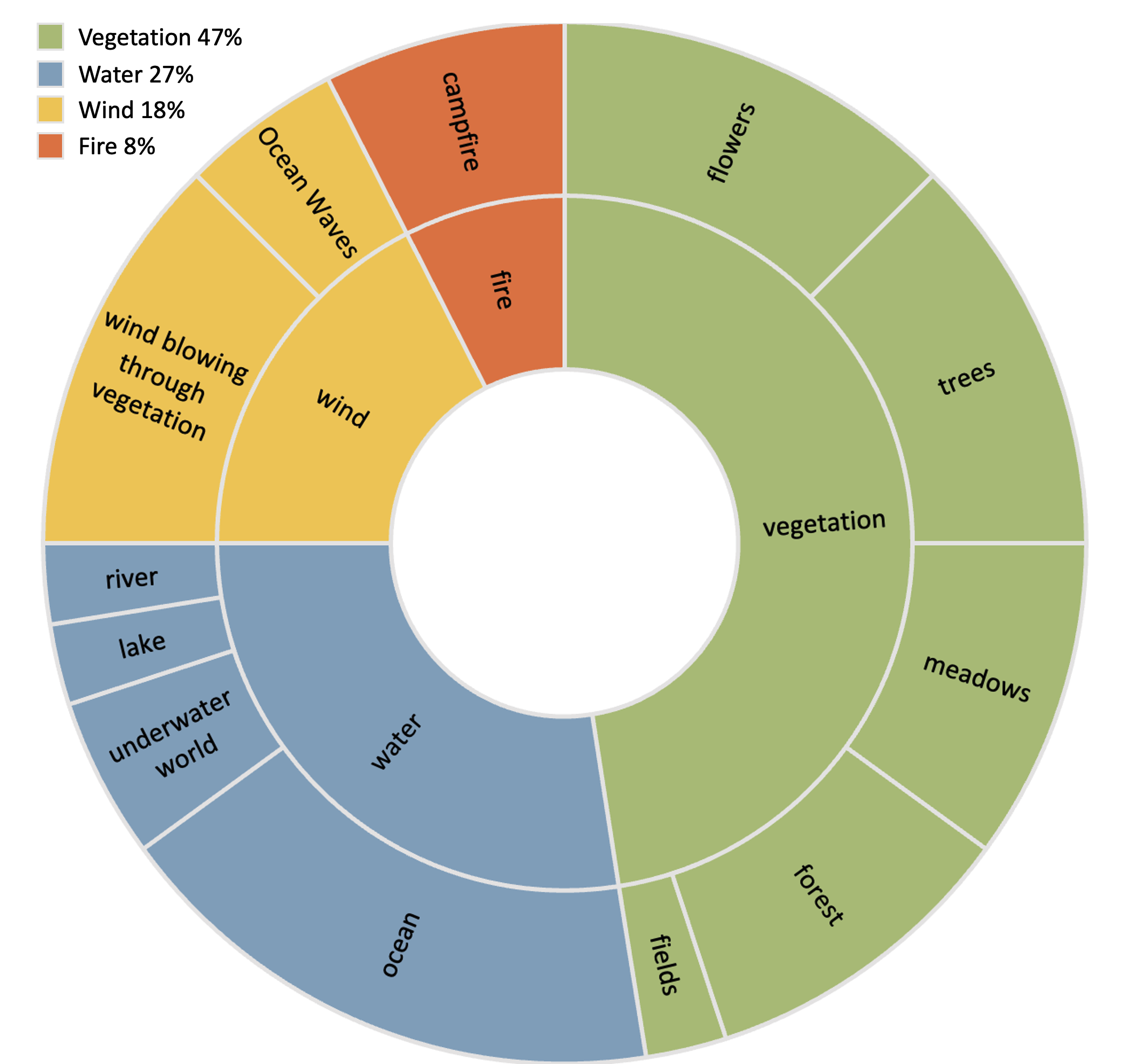}
  \caption{Virtual environment scenes categorized.}
  \label{fig:VE}
\end{figure}

~\textbf{Virtual Environments.} 
Excessive intricacy in virtual worlds might result in mental weariness and cognitive overload~\cite{kaplan2001meditation}.
Simulation of nature~\cite{calvo2014positive} and vague abstract environment~\cite{moseley2017deep} help positive meditation, which is confirmed in our review (see Figure~\ref{fig:VE}). Although the design of Ralph Moseley ~\cite{moseley2017deep} only studied abstract virtual environments as well as the work of Kazzi et al. ~\cite{kazzi2018effects} did not indicate specific virtual environments, water (ocean~\cite{kosunen2016relaworld,roo2017inner,prpa2018attending,min2020effects,andersen2017preliminary,yildirim2020efficacy,edirisooriya2019smartphone}, river~\cite{heeter2020interoceptive}, underwater world~\cite{paredes2018driving,stepanova2020jel}, lake~\cite{andersen2017preliminary}), vegetation (trees~\cite{kosunen2016relaworld,roo2017inner,patibanda2017life,mevleviouglu2021visual,feinberg2022zenvr}, flowers~\cite{kosunen2016relaworld,roo2017inner,mevleviouglu2021visual,lai2021transcent,heeter2020interoceptive}, forest~\cite{jarvela2021augmented,lai2021transcent,lee2021study,edirisooriya2019smartphone}, fields~\cite{min2020effects}, meadows~\cite{roo2017inner,mevleviouglu2021visual,andersen2017preliminary,heeter2020interoceptive}), wind (wind blowing through vegetation~\cite{kosunen2016relaworld,roo2017inner,jarvela2021augmented,mevleviouglu2021visual,yildirim2020efficacy}, wind blowing through ocean, waves~\cite{kosunen2016relaworld,edirisooriya2019smartphone}) and fire (campfire~\cite{roo2017inner,jarvela2021augmented,lee2021study}) are common features of all virtual scenes used for meditation. The abstract environment of Ralph Moseley's work~\cite{moseley2017deep} also metaphorically represents the information related to water, fire, and wind. These virtual environments all point to nature, softness, vividness, and slow movement. Although the work of Kazzi et al. ~\cite{kazzi2018effects} does not specify the specific virtual environment they used, and it shows that they used vivid and soft colors for the background of the virtual environment. 
Only the work of Madzin et al.~\cite{madzin2021re} took a different virtual environment from all the others; it built galleries as virtual scenes. 

Notably, both two studies~\cite{min2020effects, mevleviouglu2021visual} contrasted calm natural scenes with scenes of stressful stimuli. 
The stimulating scenarios they selected were cluttered and unsettling indoor and moving elevator surroundings at high heights. These scenes with a high level of detail and rapid action are more likely to produce mental weariness.

~\textbf{Virtual Objects.}
In addition to the virtual environment, the variety of virtual objects will be more abundant. Nevertheless, it is often connected to the sort of meditation, such as breath and attention.
Only the RelaWorld~\cite{kosunen2016relaworld} and DYNECOM~\cite{jarvela2021augmented} showed humanoid objects in the virtual environment; the former was an animated figure, serving as visual guide, to assist body scanning exercises. The latter were some statues that were avatars of the user in the virtual environment. It is worth noting that both designs of humanoid objects are androgynous, which was intentionally designed to have no strong identifiers and to consider all genders. 

The virtual objects or virtual environments in the six designs~\cite{roo2017inner,patibanda2017life,prpa2018attending,stepanova2020jel,jarvela2021augmented,mevleviouglu2021visual} changed in response to the participant's breathing. 
The subsequent section will describe specific methods and feedback for detecting breathing, and this section describes in detail how the virtual objects are altered with breathing. In the design of RelaWorld~\cite{kosunen2016relaworld}, with breathing, virtual objects such as clouds and the sea will move slowly, and the user exhales, and the bonfire will be enhanced. 
In the design of Lift Tree~\cite{patibanda2017life}, the user breathes in, the trunk expands, while the user breathes out; the trunk contracts. The tree will become vivid and colorful as the user practices breathing (PLB). In the design of Attending to Breath~\cite{prpa2018attending}, the user's position will rise and fall by inhaling and exhaling. And the jellyfish in the virtual underwater environment in JeL~\cite{stepanova2020jel} will respond directly to the user's breathing, and the coral will grow with the breathing. In the design of Roo et al.~\cite{roo2017inner}, breathing can control a multitude of virtual objects, such as the ripple effect of water, the movement of clouds, the flame effect of a campfire, and even the control of sound effects.
If breathing is detected, some visual effects will be activated in the virtual environment of DYNECOM~\cite{jarvela2021augmented}, such as the bar light on the bridge. The flowers in the virtual environment in the work of Mevlevioğlu et al.~\cite{mevleviouglu2021visual} will move according to the user's breathing. A small number (1/3) of these virtual objects with respiratory feedback gave only generalized feedback on the behavior of breathing (e.g., flowers or trees swaying with breathing), but most (2/3) gave different feedback on inhalation and exhalation (e.g., inhalation going up and exhalation going down). 

\begin{figure*}[h]
  \centering
  \includegraphics[width=.9\linewidth]{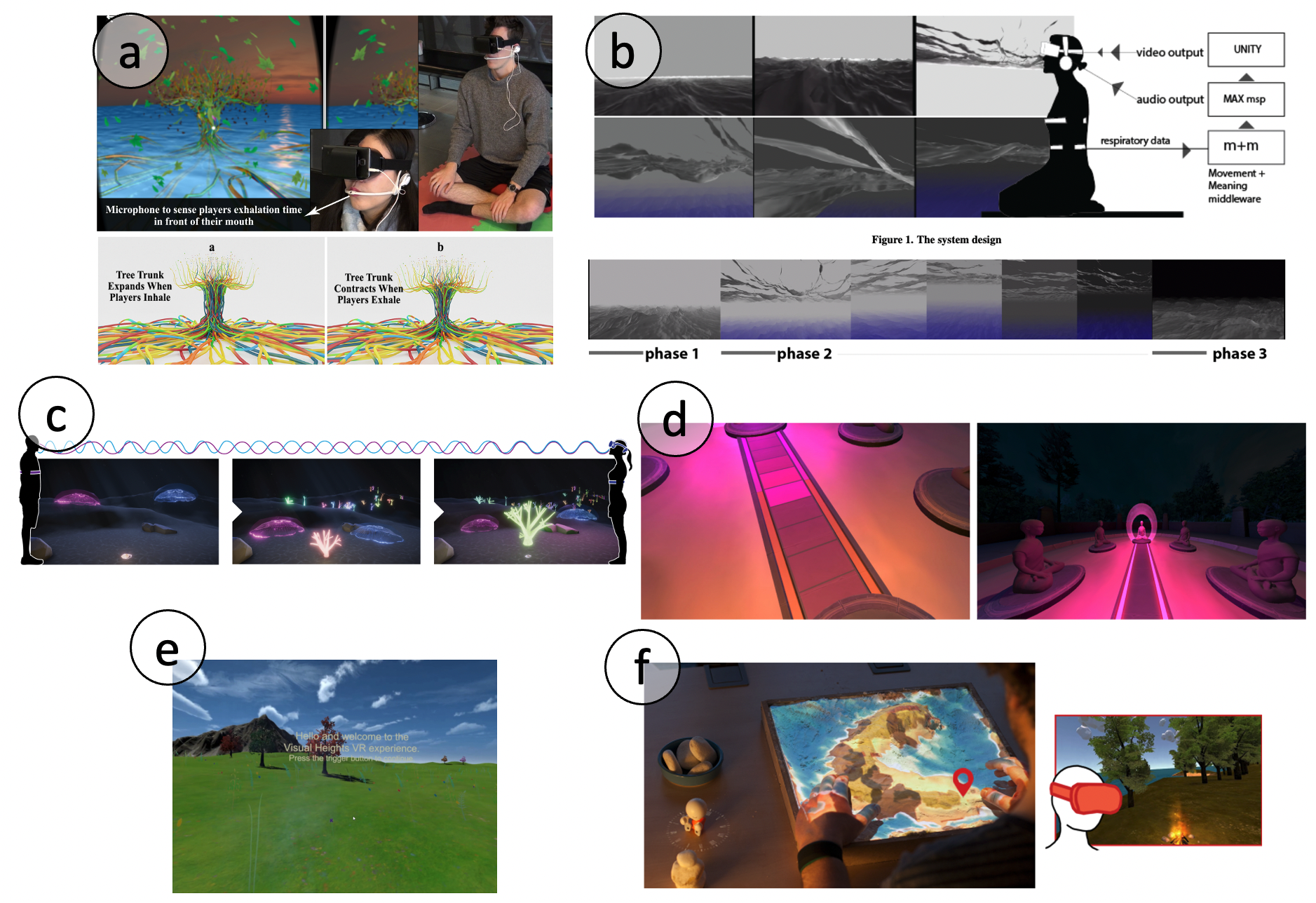}
  \caption{Example of design for breathing to interact with virtual objects; Following the rhythm of the participant's breathing, virtual objects are represented in the form of: a) tree trunks contracting and expanding~\cite{patibanda2017life}; b) people moving up and down in the virtual environment~\cite{prpa2018attending}; c) jellyfish moving and coral growing~\cite{stepanova2020jel}; d) lighting effects on bridges~\cite{jarvela2021augmented}; e) flowers and grass swaying with the wind~\cite{mevleviouglu2021visual}; f) clouds and sea moving with the wind, campfires getting stronger and sound effects changing~\cite{roo2017inner}.}
  \label{fig:breath}
\end{figure*}

Depending on the research topics, distinct studies may use distinct virtual objects.
For example, the RelaWorld \cite{kosunen2016relaworld} used a simple spherical object as the object of detection for focal meditation, while the work of Min et al.  \cite{min2020effects} used a heart-shaped symbol and a heart organ model. The design of Lee et al. \cite{lee2021study}, on the other hand, requires users to have interactive behaviors such as pruning trees. And the system changes the size of the campfire according to the user's EEG.

We found that except for the specific requirements of the study, such as the RelaWorld \cite{kosunen2016relaworld} and the design of Min et al. \cite{min2020effects}, other virtual objects that physiological feedback changes in the general virtual environment need to take comfortable objects, conform to their natural state, and have no obvious marking features. For example, the study in the work of PAREDES et al.  ~\cite{paredes2018driving} showed that people do not like to have underwater experiences with scary animals such as sharks.

~\textbf{Virtual self-representation.}
We refer to the concept in the paper of Döllinger et al. ~\cite{dollinger2021challenges}, the virtual self-representation as a specific perceptual subjective body of the user's avatar in the virtual environment, visual, auditory, or haptic. Virtual self-representation affects the perception of the body. For example, the breathing behavior in Attending to Breath~\cite{prpa2018attending} controlled the vertical position of the user's avatar in the virtual space. In the design of Driving with Fishes~\cite{paredes2018driving}, the user felt like swimming in a virtual ocean in a moving car. In addition to the visual experience, the design of Min et al.~\cite{min2020effects} also provided a haptic experience for the user. The haptic experience is achieved through physical feedback from the real world. And DYNECOM~\cite{jarvela2021augmented} adopted statues as virtual agents for users, who can also see virtual avatars of other users (the same statues) in the virtual environment. In addition to the above virtual self-representation, TranScent in Stillness~\cite{lai2021transcent} provided olfactory feedback to the user using fragrance, where the user could move around the virtual environment and smell different types of fragrances at different locations.

Except for the work in DYNECOM~\cite{jarvela2021augmented}, the rest of the reviewed work does not have a specific avatar designed for the user, who can only feel the self-identification through visual, haptic, or olfactory feedback. To the best of our knowledge, no work systematically investigates the relationship between virtual self-representation and meditation.

\subsubsection{Physiological Signals Detection}
This section focuses on the devices used to detect physiological signals in different studies and the role of detecting different physiological signals for meditation in VR.

Out of the 19 literature we reviewed, 10 studies (52.6\%) performed measurements of physiological signals, and the types of physiological data measured were electroencephalography (EEG)~\cite{kosunen2016relaworld,mevleviouglu2021visual}, heart rate~\cite{roo2017inner,min2020effects,mevleviouglu2021visual}, electrocardiogram (ECG)~\cite{kazzi2018effects}, respiratory~\cite{roo2017inner,patibanda2017life,prpa2018attending,stepanova2020jel,jarvela2021augmented}, blood pressure (finger blood pressure~\cite{kazzi2018effects} or dominant arm blood pressure~\cite{min2020effects,kazzi2018effects}), and skin conductance~\cite{min2020effects,mevleviouglu2021visual}.
 
\textbf{Equipment.} 
The review found no duplication in the devices used to detect physiological signals, and the same signal was also detected using different devices for each study. See Table~\ref{tab:equip} for a specific summary.

\begin{table*}[h]
  \caption{Summary of equipment used to measure physiological values}
    \renewcommand\arraystretch{1.5}
  \label{tab:equip}
 \resizebox{\linewidth}{!}{ 
  \begin{tabular}{p{100pt}p{400pt}}
    \toprule
    Physiological signals & Equipment \\
    \hline
    EEG & MyndPlay headband~\cite{mevleviouglu2021visual}; Muse headband~\cite{lee2021study}\\
    Heart rate & MioFuse Smartwatch~\cite{roo2017inner}; Shimmer~\cite{jarvela2021augmented}; PSL-iECG~\cite{min2020effects}\\
    ECG & Nexfin~\cite{kazzi2018effects}\\
    Respiratory & Self-made belt based on tensile sensor~\cite{roo2017inner}; Breathing+~\cite{patibanda2017life}; Thought Technology~\cite{prpa2018attending}; Biosignal Plux PZT~\cite{stepanova2020jel};  QuickAmp~\cite{jarvela2021augmented}\\
    Blood pressure & Omron M7 Intelli IT~\cite{stepanova2020jel}; sphygmoCor XCEL AtCor Medical~\cite{kazzi2018effects}\\
    Skin conductance & ALaxtha PolyG-A~\cite{stepanova2020jel}; Shimmer~\cite{mevleviouglu2021visual}\\
    \bottomrule
\end{tabular}}
\end{table*}

\textbf{Physiological feedback.}
From these reviewed studies, we found that breathing was the physiological data most susceptible to interacting with the virtual environment and providing feedback on breathing to the virtual environment or virtual objects. All five studies reviewed that detected breathing responded to breathing as physiological feedback to the virtual environment and interacted with the user. Breathing is an important component of meditation as well~\cite{hewitt2012complete}, where the user involuntarily adjusts their breathing as they observe the virtual object or environment change with their breathing in the virtual environment~\cite{patibanda2017life}.

EEG is thought to be associated with cognitive processes~\cite{klimesch1998induced}, and the author of RelaWorld~\cite{kosunen2016relaworld} argues that direct observation of the brain (recording EEG) is more meaningful than recording some peripheral physiological indicators such as heart rate or skin conductance. RelaWorld~\cite{kosunen2016relaworld} used EEG directly as physiological signal feedback, and they used the $\alpha$ band as an indicator of concentration and the $\theta$ band as an indicator of relaxation to change the transparency of the bubble in the virtual environment. The paper of Lee et al.~\cite{lee2021study} also employed EEG as physiological feedback, with virtual objects campfires changed according to the $\alpha$ band of EEG waves.

\section{Discussion}

\subsection{Meditation type and mindfulness task}

More than half of the articles do not adequately specify the meditation type or specific mindfulness job, such as the meditation type, how to execute it step-by-step, the senses engaged, and whether or not the eyes must be closed. Among publications that provide a full explanation of meditation content, concentration breath meditation is the most prevalent form. The Kasina, Compassion meditation, and Stability meditation are three specific meditations explored in our collected publications, and their mediators or relevant design concerns are centered on the virtual object, empathy, and the sensation of stability~\cite{moseley2017deep,jarvela2021augmented,heeter2020interoceptive}.

Meditation has a lengthy history and several variations due to its Buddhist origins and introduction to psychiatric treatment. Although there are ostensibly basic prerequisites for meditation, various types/tasks/techniques may be tailored to individual meditators or design factors. The relaxation response, developed by physiologist Herbert Benson and based on transcendental meditation, comprises four key components: a peaceful atmosphere, a comfortable posture, a passive attitude, and a mental device. The passive attitude is the lack of judgment that allows any notion to ``pass through.'' The mental device involves concentrating on a word, sound, phrase, or rhythmic breath~\cite{benson1975relaxation, dienstfrey1991mind}.

A successor, i.e., another psychologist --  Stephen Kaplan, identified fundamental connections between attention restoration theory and meditation. Entering a restorative setting will restore one's direct attention, which has limited capacity, according to the notion of attention restoration. There are four levels of properties of this restorative environment: being away means the environment is unique and allows the person to escape the stresses of daily life; fascination means the visual pattern in the environment can hold the person's attention; extent means the environment has the scope and coherence that allow the person to remain engaged, and; compatibility means the environment is conducive to the activity that the individual desires to pursue. These four qualities of environments are more concerned with the outcomes of the person-environment interaction than with defining the environment itself~\cite{kaplan2001meditation}. Meditation has comparable elements. Take breath meditation as an example; its processes include sitting, shutting one's eyes, taking deep breaths (being away), concentrating on the breath, and counting breaths (fascination and extent). The distinction is that meditation emphasizes the active participation of the individual, which may need training and abilities~\cite{kaplan2001meditation}.

Meditation and mindfulness have not attained a general agreement in psychology or related fields. Bishop~\cite{bishop2004mindfulness} provided an operational definition of meditation that consists of two elements. One element is the self-regulation of attention to sustain focus on the current experience and permit awareness of mental events occurring in the present moment. The unique attitude toward one's experience in the present moment, characterized by openness, curiosity, and acceptance, is another component.

In summary, the common characteristics mentioned in numerous disclosures include the separation of daily stressful life, the reduction of information stimuli requiring high-cognitive resources, the concentration on specific objects (interoception such as one's breath or body parts' feelings), the awareness of the present moment, and the maintenance of an attitude of openness, acceptance, and non-judgment.

\subsection{Design considerations and impacting variables}
According to our results, most articles don't have a detailed description of the meditation tasks. They only achieve being away from the daily stressful life and hold one's attention to some extent via VR's immersive experience. Reviewing our design consideration and impacting variables coding results, one-third of articles aimed only to improve the presence of users, which haven't released the potential of VR. In contrast, some other works are very informative. The InnerGarden considered the enhance the user's acceptance attitude by an irreversible procedure design ~\cite{roo2017inner}. 

The Kasina meditation method corresponds extremely well with the production of virtual scenes and objects~\cite{moseley2017deep}. The Attending to Breath system uses the underwater situation and biofeedback visual signals in the virtual world to improve people's focus and breath awareness~\cite{prpa2018attending}. In addition to articles concentrating on VR technology, certain mobile phone-based works are equally informative. For instance, \textit{Pause}\footnote{\url{https://www.pauseapp.com/}} is a smartphone application for interactive meditation, and its design was informed by relaxation response and attention recovery theory.

\subsection{Virtual Reality and related technology}
Virtual environments are endlessly possible and can be created as needed to make the meditation environment more interesting. However, content that is too interesting for positive thinking can distract the user's attention, which is contrary to the design goal of the system~\cite{roo2017inner}. 
Future researchers or designers tasked with creating virtual environments for meditation should consider limiting the degree of user engagement. Users will focus more on the physical components of the virtual world than on its engaging content.

According to the survey results, individuals dislike virtual items that may induce fear~\cite{paredes2018driving}, as well as virtual settings with a specific height~\cite{mevleviouglu2021visual} or distracting clutter~\cite{min2020effects}. Fast-moving scenes may induce motion sickness in users~\cite{prpa2018attending,paredes2018driving,heeter2020interoceptive}. Therefore, the design of the virtual environment should be natural and pleasant, and the user's posture inside the virtual environment should stay fixed or move slowly. Designers might think about adding vertical movement, which is thought to cause less motion sickness than horizontal movement~\cite{prpa2018attending}.

Prior research provides real-time input of the user's physiological data to the virtual world, which is then represented by specific virtual objects. It should be noted that too immediate reactions to physiological changes might increase user anxiety since such quick responses can divert users' attention and make it harder for them to grasp the relationship between mental state and environmental adaption~\cite{kosunen2016relaworld}. Furthermore, physiological data like heartbeats, which are fast and uncontrollable by the user, are unsuitable for immediate feedback~\cite{roo2017inner}. Consequently, breathing is the optimal data for physiological feedback and engagement with a virtual world. It is sluggish and user-controllable, making it the most popular approach in the evaluated literature. Researchers may explore employing a photoplethysmogram (PPG) instead of an electrocardiogram (ECG) to record the heart rate for more flexibility~\cite{mevleviouglu2021visual} when capturing physiological data for further analysis.

\subsection{Methodological Limitations}
In this work, we conducted a systematic literature review in two databases (i.e., ACM and IEEE). As meditation, or mindfulness, is closely relevant to medical, health, and psychology, other databases may provide insightful studies and findings, for example, Pub-Med, ScienceDirect, and Springer. 
Furthermore, we only used keyword combinations of synonyms for meditation and VR. However, some studies may use other terminologies like breath training, deep breathing, respiration practice, etc. We might miss these insights in these papers. 
Due to the small number of available publications in the selected database, we did not adopt an assessment of the quality of a paper before coding it, so the varying quality of collected publications might contribute equally to our results, which may lead to some bias in the results. 

\section{Opportunities \& Challenges for Future Research}

Through the systematic literature review, we have identified five opportunities to further leverage VR to improve meditation experiences. 

First, researchers must go further into the \textbf{specific meditation type or mindfulness activity} to determine the processes, mediators, and objectives that are compatible with the appropriate technology and maximize the potential of VR and similar technologies. For instance, virtual reality (VR) settings and physiological detection might be used to improve interoceptive awareness. What kinds of virtual surroundings or activities are most conducive to fostering acceptance and non-judgment in individuals? And, what activities might govern a person's focus on the present? How about combining mindfulness with other artistic endeavors? Chinese calligraphy might be a suitable starting point, since it is known as a mental development exercise (in Chinese: Xiu Shen Yang Xing). When practicing Chinese calligraphy, a person pays close attention to hand movement, the writing brush, and the writing style on the paper.

Second, future research could examine \textbf{self-representation and environment representation} in VR and create a transcendental experience for users, enabling them to engage in profound introspection and life reflection, this is possible with the rapid development and vast potential of 3D modeling technology. In addition to the self-representation of the figure, according to our survey, at this point in time, the majority of research has used physiological feedback to move objects in virtual settings, such as trees or clouds. Only one of the experiments~\cite{jarvela2021augmented} we analyzed included partial physiological data input into the virtual representation of the user. If physiological data may be immediately responded to by virtual avatars, is it feasible to increase users' sense of self? This merits additional investigation.

Third, the two research opportunities mentioned above both rest on the exploration of two senses, visual and auditory; is it worth exploring \textbf{multisensory} stimulation and interaction in virtual reality meditation? In addition to the visual and auditory stimuli that would have been used in the majority of research, haptic and olfactory sensations seem to have a good influence on meditation~\cite{roo2017inner,min2020effects,lai2021transcent}. However, this is not supported by all investigations. In the future, researchers might try out alternative sensory stimuli to see if various stimuli help people pay attention and think about things.

Fourth, future researchers may need to be more concerned with the \textbf{types of interactions} meditators have in virtual reality. Interaction techniques in virtual reality play a key role in enhancing presence and immersion~\cite{slater1994body,kosunen2016relaworld}, but rich engagement with virtual environments can also contribute to distraction~\cite{roo2017inner}. The question of whether virtual environments must be minimalist remains contentious~\cite{terzimehic2019review}. It is also debatable if VR is, in fact, helpful for meditation. Nevertheless, it is obvious that VR technology has expanded the opportunities for meditation, particularly after COVID-19, and that the demand for digitally aided meditation has risen~\cite{lai2021transcent}. Future studies will face the challenge of determining how to construct virtual worlds that provide the necessary immersion and sense of concentration for meditation.

Fifth, according to the findings of our analysis, the great majority of studies (14/19) concentrated only on individual meditation practices and disregarded the \textbf{social aspects} of VR technology. A few studies~\cite{jarvela2021augmented, stepanova2020jel} have investigated group meditation or social meditation. Meditation is an activity that demands attention to oneself and self-reflection, but in modern culture, communal healing and calming actions are also quite significant~\cite{kohrt2020we}. The instruments of digital technology should be able to promote communal introspection, emotional contagion, and empathy. Likewise, VR technology may serve as a technological medium. In a virtual world, the user's visual, aural, and tactile senses may be reproduced. Future research may concentrate on group mindfulness to remedy this gap.

\section{Conclusion}

Meditation, with origins in Buddhism, is introduced into psychology as a therapy approach to relieve stress and improve overall human well-being. Existing studies provide evidence that meditation in VR is beneficial for people. We conducted this review to understand the state-of-the-art in this area and get insights into how VR and related technology can assist meditation. We examined the ACM and IEEE databases and finally included 19 articles. Our analysis structure covers three themes: meditation or mindfulness task, design considerations (crucial factors impacting meditation and enhanced by technology), and VR and related technology.

Our results show that many (N=10) studies do not clearly define the meditation type in their research. A portion of the studies (N=6) conducted breath meditation Or included breath as part of a meditation activity, and the other studies (N=6) conducted Kasina, compassion meditation, etc. Part of the studies (N=7) only aimed to enhance users' presence by improving the immersion to different extents; some other articles focused on improving awareness of breath besides presence; a small portion of articles (N=3) explored interoception and attention regulation. In terms of VR equipment, All studies (N=19) used HMD displays (mainly Oculus and HTC Vive products). Most studies (N=17) involved nature-related environments like wind, the ocean, forests, and campfires. The virtual objects were closely related to the specific meditation type and contained plants, marine animals, and ocean waves. Only a few papers (N=5) explored self-representation in VR, and users felt their existence in VR through multiple senses. More than half of the studies (N=10) adopted physiological detection technologies, and EEG and respiration are treated as the most widely used indicators. Other indicators include blood pressure, heart rate, and skin conductance.

Our review results indicate that the current works have a profound accumulation of utilizing VR immersion to improve users' presence (letting them be away from their daily stressful life). Some of these works are insightful in combining physiological detection with interactive virtual objects. However, meditation contains many types and adaptations other than breath meditation. VR, 3D modeling, and physiological detection also include ample space to explore, and it is not convincing that the most advanced rendering technology, the most comprehensive multi-sensory stimulation, and the richest interaction will achieve the best meditation effect. For future research, we should aim to determine the processes, mediators, and objectives of the specific meditation type or mindfulness activity to integrate the appropriate technologies and maximize the potential.  


\bibliographystyle{ACM-Reference-Format}
\bibliography{mybib}


\end{sloppypar}
\end{document}